\documentstyle[11pt,a4]{article}
\begin{document}

\section{Introduction}

\indent

With the advent of very energetic hadronic colliders like LHC and SSC,
whose aim is to have access to the mechanism of electroweak symmetry
breaking, a precise estimate of inclusive production of neutral clusters
is mandatory to pin down signals due to Higgs particle or new
physics \cite{altap}. Perturbative QCD is the appropriate framework to
perform such a calculation provided large transfer momenta
are involved. For this purpose leading order (hereafter denoted as LO )
 predictions - based on
evaluations of partonic cross sections at tree level and evolution of
structure and fragmentation functions at one loop level
- are too rough. A consistent calculation at next to leading order
(hereafter denoted as NLO) needs a NLO evaluation of
parton-parton subprocesses which has been performed by our
group \cite{acgg}  a few years
ago utilizing existing results on the O($\alpha_s^3$)
matrix element \cite{elliss} and two loop evolved structure and fragmentation
functions. Various
sets of structure functions based on a NLO analysis using deep inelastic
lepton nucleon scattering data, Drell-Yan production and direct photon
production at hadronic
accelerators are available in the literature \cite{mt,mrs,abf}.

\indent

Up to now such an analysis has not been performed for pions
fragmentation functions since the only available derivation at NLO
has been done for heavy quarks \cite{melna}.
The parametrizations presently available \cite{bep}
are based on a LO analysis of rather old $e^+ e^-$ data and semi
inclusive deep inelastic muon nucleon scattering. Our aim is to
perform a complete NLO evaluation of neutral pions inclusive
production from hadronic colliders in order to estimate, as precisely
as possible, the
$\pi^0$ rates at LHC  and SSC. The first
step will consist in performing an extraction of pions fragmentation
functions at NLO using $e^+ e^-$ data and hadronic data on one
particle inclusive production. This will be done using three different
methods. The first one, which
does not exactly correspond to a NLO analysis but rather to an improved LO
approximation, is based on the Monte-Carlo simulator HERWIG \cite{her},
 whereas the second and the third ones
are obtained from a two loop evaluation of
evolution kernels previously computed \cite{altpa,furm1}
together with NLO calculation of one hadron inclusive
production from $e^+ e^-$ \cite{alta1} and hadronic colliders
\cite{acgg} using respectively natural and optimized scales.
As we will see it is not possible to extract an unique set
of fragmentation functions which fits $e^+ e^-$ data around
$\sqrt{S}=30$ GeV \cite{cello1,cello2,tasso,tpc,jade}
and hadronic data from fixed target domain \cite{wa70,e706}
to collider range \cite{afs,kou,ua2}.
We will therefore take different sets corresponding to different
hypotheses on input fragmentation functions.

\indent

The paper is organized as follows. We recall the expressions of one
particle inclusive production from $e^+ e^-$ and hadronic collisions and
also give the evolution equations for fragmentation functions in
section 2. Then in section 3 we discuss the extraction of various sets of
fragmentation functions, first from HERWIG simulation and after through an
exact NLO derivation.
Predictions at LHC using the different sets previously
obtained are displayed in section 4 together with a discussion of the resulting
theoretical uncertainty. We give our conclusions in section 5.

\section{One particle inclusive production at next-to-leading order}

\indent

Let us consider the inclusive production of a hadron $H$ via the
generic reaction $A+B \rightarrow H$ where $A$ and $B$ stand for hadrons
and/or leptons. The cross-section can be written as a convolution
of the fragmentation functions $D_l^H(z,M_f^2)$ with the partonic
cross-section:
\begin{equation}
E_{H} \; \frac{d\sigma_{A+B \rightarrow H}}{d^3 \vec{P}_{H}}
 =  \sum_{l} \; \int_{z_H}^{1} \frac{dz}{z^2} \; D^{H}_{l}(z,
M^2_f) \; E_l \: \frac{d\sigma_{A+B \rightarrow l}}{d^3 \vec{P}_{l}}
(\frac{z_H}{z}, \theta, \alpha_s(\mu^2), M^2_f,\cdots) ,
\label{meq}
\end{equation}
where $z_H$ is the reduced energy of the hadron $H$:
$z_H=2 E_H / \sqrt S$ and
$\theta$ is the scattered angle of the parton l.
The inclusive production of the parton l via the reaction
$A+B \rightarrow l$
has the following perturbative development:
\begin{equation}
E_l \; \frac{d\sigma_{A+B \rightarrow l}}{d^3 \vec{P}_{l}}
(\frac{z_H}{z}, \theta, \alpha_s(\mu^2), M^2_f,\cdots ) =
\sigma^0_{A+B \rightarrow l}(\frac{z_H}{z},\theta) + \frac{\alpha_s
(\mu^2)}
{2 \pi} \; \sigma^1_{A+B \rightarrow l}(\frac{z_H}{z},\theta,M^2_f) +
\cdots .
\end{equation}
Finally $D^H_l(z,M^2_f)$ represents the number of hadrons H
inside the parton l carrying the fraction of impulsion z from H,
evolved at the scale $M^2_f$.
These fragmentation functions satisfy Altarelli-Parisi type
evolution equations:
\begin{eqnarray}
\frac{\partial D_q^{H}(z,M^2_f)}{\partial \ln(M^2_f)} & = & \frac{
\alpha_s(M^2_f)}{2 \pi} \int_{z}^{1} \frac{dy}{y} \left[ P^T_{qq}
(y,\alpha_s(M^2_f)) D^{H}_q(\frac{z}{y},M^2_f) \right. \nonumber \\
& & \mbox{} + \left. P^T_{gq}
(y,\alpha_s(M^2_f)) D^{H}_g(\frac{z}{y},M^2_f) \right]  \\
\frac{\partial D_g^{H}(z,M^2_f)}{\partial \ln(M^2_f)} & = & \frac{
\alpha_s(M^2_f)}{2 \pi} \int_{z}^{1} \frac{dy}{y} \left[ P^T_{qg}
(y,\alpha_s(M^2_f)) D^{H}_q(\frac{z}{y},M^2_f) \right. \nonumber \\
& & \mbox{} + \left.  P^T_{gg}
(y,\alpha_s(M^2_f)) D^{H}_g(\frac{z}{y},M^2_f) \right] .
\end{eqnarray}
The evolution kernels have the perturbative development:
\[ P^T_{ij}(x,\alpha_s(M^2_f)) = P^0_{ij}(x) + \frac{\alpha_s(M^2_f)}
   {2 \pi} P^{T1}_{ij}(x) + \cdots . \]
In the following, we will drop the superscript index T.
The Altarelli-Parisi kernels have been computed up to two loops
order by Curci, Furmanski and Petronzio \cite{furm1}.
In the LO approximation one keeps only
the first order in the perturbative development of the partonic
cross-section and in the evolution kernels
whereas at NLO one keeps the
first and second terms in the perturbative expansion for both
partonic cross-section and evolution kernels.
We can split these fragmentation functions into a non-singlet
and a singlet part:
\begin{eqnarray}
D_i^-(z,M_f^2) & \equiv & \frac{1}{2} \left( D_{q_i}^H(z,M_f^2)-
 D_{\bar{q}_i}^H(z,M_f^2) \right)   \\
D_i^+(z,M_f^2) & \equiv & \frac{1}{2} \left( D_{q_i}^H(z,M_f^2)+
 D_{\bar{q}_i}^H(z,M_f^2) \right) - \frac{1}{2 N_f} D_S(z,M_f^2) \\
D_S(z,M_f^2) & \equiv & \sum_{i=1}^{N_f} \left( D_{q_i}^H(z,M_f^2)+
 D_{\bar{q}_i}^H(z,M_f^2) \right) .
\end{eqnarray}
In the evolution equations
the singlet part $D_S$ is coupled to the gluon fragmentation function
whereas the
non-singlet parts $D^-$ and $D^+$ are decoupled.
Note a misprint in ref.~\cite{furm1} and the correspondence
with our notation:
\[ \begin{array}{ccc}
P^i_{gg} & = & P_{GG} \\
P^i_{qg} & = & P_{GF}/(2 N_f) \\
P^i_{gq} & = & 2 N_f \: P_{FG} \\
P^i_{qq} & = & P_{FF} .
\end{array} \]
\\
Once input fragmentation functions have been specified  at some reference
scale $M_{f0}$ the evolution equations are solved using an inverse Mellin
transform technique.
Let us consider now in detail the partonic cross-sections.

\subsection{$e^+ e^- \rightarrow \pi^0$}
\indent

The partonic cross sections from $e^+e^-$ collisions read at
next-to-leading order:
\begin{eqnarray}
\lefteqn{E_{q_i} \: \frac{d\sigma_{e^+ + e^- \rightarrow q_i}}{d^3
\vec{P}_{q_i}} (y,\theta,\alpha_s(\mu^2),M^2_f) = } \nonumber \\
&  & \frac{6 \: \sigma_0}{\pi Q^2 y}
\; e_i^2 \left\{ \frac{3}{8} (1+\cos^2 \theta) \left[
\delta(1-y) + \frac{\alpha_s(\mu^2)}{2 \pi} \left(
P^0_{qq}(y) \ln\!\left( \frac{Q^2}{M^2_f} \right) + K_q^T(y) \right)
\right] \right. \nonumber \\
& & \mbox{} + \left. \frac{3}{4} (1-\cos^2 \theta)
\frac{\alpha_s(\mu^2)}{2 \pi} K^L_q(y) \right\} \\
\lefteqn{E_g \: \frac{d\sigma_{e^+ + e^- \rightarrow g}}{d^3 \vec{P}_{g}}
(y,\theta,\alpha_s(\mu^2),M_f^2) = } \nonumber \\
&  & \frac{12 \: \sigma_0}{\pi Q^2 y}
\sum_{i=u,d,s,c,...} e_i^2
\left\{ \frac{3}{8} (1+\cos^2 \theta)
\left[ \frac{\alpha_s(\mu^2)}{2 \pi} \left(
P^0_{gq}(y) \ln\!\left( \frac{Q^2}{M^2_f} \right) + K_g^T(y) \right)
\right] \right. \nonumber \\
& & \mbox{} + \left. \frac{3}{4} (1-\cos^2 \theta)
\frac{\alpha_s(\mu^2)}{2 \pi} K^L_g(y) \right\} ,
\end{eqnarray}
where $\sigma_0$ is the usual point like cross-section
\[ \sigma_0 = \frac{4 \pi \alpha^2}{3 Q^2} , \]
$\alpha$ is the QED coupling constant and $Q^2$ is the invariant
mass of the $e^+ e^-$ pair.
The functions $K^T_q$, $K^L_q$, $K^T_g$ and $K^L_g$ have
been extracted from the
reference \cite{alta1} (see also \cite{others}).
\begin{eqnarray}
K^T_q(x) & = & C_F \; \left\{ \frac{3}{2} (1-x) - \frac{3}{2}
\frac{1}{(1-x)_+} + 2 \frac{1+x^2}{1-x} \ln(x) \right. \nonumber \\
& & \mbox{} + \left. (1+x^2) \left( \frac{\ln(1-x)}{1-x} \right)_+
+ \left(
\frac{2 \pi^2}{3}-\frac{9}{2} \right) \delta(1-x) \right\} , \\
K^T_g(x) & = & C_F \; \left\{ \frac{1+(1-x)^2}{x} \left( \ln(1-x) + 2
\ln(x) \right) - 2 \frac{1-x}{x} \right\} \\
K^L_q(x) & = & C_F  \\
K^L_g(x) & = & 2 \; C_F \; \frac{1-x}{x} .
\end{eqnarray}
In the above equations two scales are involved:
the renormalization scale $\mu$ at which the
running coupling constant $\alpha_s$ is evaluated and
the fragmentation scale $M_f$ at
which fragmentation functions are evolved. The choice for these
scales is rather arbitrary.
Note that for every $y$, $K^T_g(y)$ is negative, so the choice
$M_f^2=Q^2$
leads to a negative contribution to the partonic cross-section
$E_g \; d \sigma_{e^+ e^-\rightarrow g}/ d^3 \vec{P}_g$.

\indent

The running coupling of QCD $\alpha_s$ is defined at the next-to-leading
logarithm approximation by the approximate analytical formula:
\begin{equation}
\alpha_s (\mu^2) = \frac{1}{b \ln (\mu^2/\Lambda^2)} \; \left[
1 - \frac{b'}{b} \: \frac{\ln \ln (\mu^2/\Lambda^2)}{\ln (\mu^2
/\Lambda^2)} \right].
\label{apalfa}
\end{equation}
In section 3 we will also use for $\alpha_s$
the numerical solution of the equation:
\begin{equation}
\frac{1}{\alpha_s(\mu^2)} + b'\: \ln\!\left( \frac{b' \alpha_s(\mu^2)}
{1 + b' \alpha_s(\mu^2)} \right) = b \: \ln\!\left( \frac{\mu^2}
{\Lambda^2} \right) ,
\label{trualfa}
\end{equation}
with:
\[ b = \frac{33-2N_f}{12 \pi}, \; \; b'= \frac{153 -19 N_f}{24 \pi^2}
, \]
which is more appropriate than eq.(14) for small scales
$\mu$. Indeed for
large $\mu$ the two definitions agree but for small $\mu$ they can
differ by more than 20 \%.

\subsection{$p \; p  \rightarrow \pi^0$}
\indent

The partonic cross-sections for hadronic collisions are given by
\cite{acgg}:
\begin{eqnarray}
E_l \frac{d\sigma_{p+p \rightarrow l}}{d^3 \vec{P}_{l}}
(y,\theta,\alpha_s(\mu^2),M_f^2) & = &
\frac{1}{\pi S} \sum_{i,j} \int_{V W}^{V} \frac{dv}{1-v}
\int_{V W/v}^{1} \frac{dw}{w} \nonumber \\
& & \mbox{} \times \left[ F^p_i(x_1,M^2)
F^p_j(x_2,M^2) \left( \frac{1}{v} \left( \frac{d \sigma^0}{dv}
\right)_{i j \rightarrow l} (s,v) \delta(1-w) \right. \right. \nonumber
\\
& & \left. \left. + \frac{\alpha_s(\mu^2)}{2 \pi} K_{i j \rightarrow l}
(s,v,w;\mu^2;M^2,M_f^2) \right) + (x_1 \leftrightarrow x_2) \right] .
\end{eqnarray}
The variables V, W are defined by
\[ V = 1- \frac{y}{2} (1-\cos \theta), \; \; W= \frac{y (1+\cos \theta)}
{2 - y (1-\cos \theta)} , \]
and we also have
\[ x_1= \frac{VW}{vw} , \; \; x_2  = \frac{1-V}{1-v} \]
and $s=x_1 x_2 S$.
At NLO sixteen subprocesses contribute to the cross-section.
The terms $\sigma^0$ correspond to the lowest order $2 \rightarrow 2$
parton scattering subprocesses whereas the terms $K$ contain
the one loop corrections to these subprocesses.
In the hadronic case, we have three scales: the renormalization
scale $\mu$, the factorization scale for the initial state $M$ (
the scale of the distribution functions) and the factorization
scale for final state $M_f$
(the scale of the fragmentation functions).
Schematically, the hadronic cross-section can be written as:
\begin{eqnarray}
E_{\pi^0} \frac{d\sigma_{p+p \rightarrow \pi^0}}{d^3 \vec{P}_{\pi^0}}
& = & \alpha_s^2 (\mu^2)
A + \alpha_s^3(\mu^2) \left[ 2 b A \ln\!\left( \frac{\mu^2}{\Lambda^2}
\right) + B \ln\!\left( \frac{M^2}{\Lambda^2}
\right) \right. \nonumber \\
 & & \left. + C \ln\!\left(\frac{M_f^2}{\Lambda^2}\right) + D \right] .
\end{eqnarray}
We show explicitly the dependence of the hadronic cross-section upon
the three scales $\mu$, $M$ and $M_f$. The four functions A, B, C and
D depend on the scales $M$ and $M_f$ via the structure and fragmentation
functions. In addition, A, B and C are scheme independent.
We always use the $\overline{MS}$ scheme for final factorization
whereas the
initial factorization scheme is fixed by the set of structure functions
used.

\indent

Let us discuss now the partonic cross-sections.
In order to determine the kinematical region where each partonic reaction
dominates we have plotted in figures 1a, 1b, 1c and 1d
the partonic cross-sections
$ E_l \; d\sigma_{p+p \rightarrow l} / d^3 \vec{P}_{l}$
for $l=g$, $u+ \overline{u}+d+\overline{d}$,
 $s+ \overline{s} + c + \overline{c}$
against $P_t$ at the leading log level for various center of mass energies.
We think it is meaningless to use next-to-leading formulae
since the dependence on $\ln(M_f^2)$ is not balanced.
We have used ABFOW structure functions \cite{abf}.
We see that for the low center-of-mass energy
experiments WA70 \cite{wa70} ($\sqrt S= 23$ GeV)
and E706 \cite{e706} ($\sqrt S= 31$ GeV)
the gluon and the valence quarks contributions
are of the same order at low $P_{tl}$, whereas
when $P_{tl}$ becomes larger, the valence
quarks dominate.
For ISR experiments \cite{afs},\cite{kou} ($\sqrt S= 63$ GeV) the glue
contribution
dominates up to $P_{tl} \simeq 10$ GeV. For the UA2 experiment \cite{ua2}, when
the pseudo rapidity $\eta=1.4$, the glue contribution is important
up to $P_{tl} \simeq 35$ GeV. Finally for LHC, in the $P_{tl}$ range
between 30 and 1000 GeV the glue contribution represents (60 - 80) \%
of the partonic cross-section. In all cases the "sea" contribution
(s,c) is always negligible.

\indent

In order to estimate the z range we are sensitive to we will study in table I
the integrand of eq.~(\ref{meq}), i.e.:
\begin{equation}
<z> = \frac{
 \int \frac{dz}{z} \sum_{l} D_l^{\pi^0}(z,M_f^2)
 \; E_l \: \frac{d\sigma_{p+p \rightarrow l}}{d^3 \vec{P}_{l}} }
{ \int \frac{dz}{z^2} \sum_{l} D_l^{\pi^0}(z,M_f^2)
 \; E_l \: \frac{d\sigma_{p+p \rightarrow l}}{d^3 \vec{P}_{l}} }
\end{equation}
with z varying between $2 E_{\pi^0} / \sqrt S$ and 1. Note that
the partonic cross-sections reach their maximum for z = 1 while
the fragmentation functions decrease with z. As we can
infer from Table I the large z region is kinematically favored.
We have used set I of fragmentation functions which will be discussed later.
\indent

As already mentioned the fragmentation functions are known less
accurately
than the structure functions. Up to now they have been extracted at LO
\cite{bep} from
$e^+e^-$ annihilation and semi inclusive deep inelastic muon nucleon
scattering. No extraction from hadronic colliders data has been
performed so far.
In the following we will carry out an extraction of $\pi^0$ fragmentation
functions at NLO accuracy, using three different approaches.

\section{Extraction of $\pi^0$ fragmentation functions.}

\subsection{Selection of experimental data.}
\indent

We first discuss the experimental data we will use to extract the
 $\pi^0$ fragmentation functions. We first consider $e^+e^-$ collisions.
The JADE collaboration \cite{jade}
has published data at $\sqrt{S}=14,~22.5$ and $34.4$ GeV.
We use the data at $34.4$ GeV, covering mainly the low $z_H$ range
(up to $z_H=0.209$).
Data from the TPC collaboration \cite{tpc}
at $\sqrt{S}=29$ GeV are given as
$\frac{1}{\sigma_{had}} \frac{d\sigma}{\beta dz_H}$, therefore a value of
R=4.00 is assumed to bring them to the usual form
$\frac{S}{\beta} \frac{d\sigma}{dz_H}$.
Data from the TASSO collaboration \cite{tasso}
at $\sqrt{S}=34.6$ GeV extend up to $z_H=0.728$.
The broadest $z_H$ range is covered by data from the CELLO
collaboration, extending from $z_H=0.049$
to $z_H=0.919$ at $\sqrt{S}=35$ GeV
\cite{cello2} and from $z_H=0.094$
to $z_H=0.847$ at $\sqrt{S}=22$ GeV \cite{cello1}.
Data from experiments at DORIS are not used, as hardly
any point survive with the cut on the lower energy
of the $\pi^0$ at 2 GeV. Data obtained at LEP
are for the moment not constraining.
However, cross checks have been performed with the
2 points surviving the cut of data from
the Argus collaboration \cite{argus} at $\sqrt{S}=10$ GeV
and the 4 points
from the L3 collaboration \cite{l3} at $\sqrt{S}=91$ GeV.

\indent

Let us discuss now experimental data from hadronic colliders.
Data in hadronic reactions have been
selected for this study taking into account statistical
and systematic accuracy.
Whenever possible, reconstructed $\pi^0$
are preferred.
For SPS fixed target energies, the available data in $pp$ reactions
are in reasonable agreement and we will use the data
in the central rapidity range at $\sqrt{S}=23$ GeV, from the
WA70 collaboration \cite{wa70}. The FNAL fixed target range
overlaps with the lower ISR energy range. The recent
data at $\sqrt S =31$ GeV  from
$p Be$ reactions obtained by the E706 collaboration \cite{e706}
are in agreement with some of the ISR results.
Resolved $\pi^0$ at $\sqrt S =62.8$ GeV taken from table 5 ( more precisely
data corresponding to the super-retracted geometry) of
Kourkoumelis et al.
\cite{kou} are used. They
will be compared with other data available at this energy.
We will use also the more recent data from the AFS collaboration
\cite{afs}, which however show a different $P_t$ dependence.
At collider energies, the latest data from the UA2 experiment
at $\sqrt S =630$ GeV with average pseudo rapidity $\eta=1.4$
will be used \cite{ua2}. Cross checks have been made with data
at $\sqrt S =540$ GeV with average pseudo rapidity $\eta=0$
although $\pi^0$ are not disentangled from direct photons.

\subsection{Fragmentation functions from HERWIG.}
\indent

We first consider the $\pi^0$ inclusive production in $e^+e^-$
annihilation at $M_{f0}=\sqrt S=30$ GeV, as simulated by the Monte Carlo
generator HERWIG. As well known, this event generator includes the QCD
parton shower to leading and next to leading accuracy - in particular
the kinematical corrections due to the phase space boundaries are
summed up to all orders - as well as the
hadronisation of the color singlet clusters into the physical particles.
Furthermore HERWIG has been shown \cite{herlep} to describe with good
accuracy the observed features of PETRA and LEP  data.
Then we will use the $\pi^0$ distribution
generated by each quark flavor which originates from the
photonic vertex, as a
realistic description of the quark fragmentation into $\pi^0$.
Owing to the
symmetry of quarks and antiquarks fragmenting into $\pi^0$
we extract the quark fragmentation functions from:
\begin{equation}
{d\sigma_{e^+e^-} \rightarrow \pi^0 \over dz_H} (z_H,M_{f0}^2) \sim
 6 \sigma_0 \sum_q e_q^2 D_q^{\pi^0} (z_H, M_{f0}^2),
\label{her1}
\end{equation}
where the pointlike cross section $\sigma_0$ has been previously defined.
The reaction $ e^+e^- \rightarrow \pi^0+ X $ has been therefore
decomposed
into each contribution $e^+e^- \rightarrow u \bar u$, $d \bar d$,
$s \bar s$, $c \bar c$ and $b \bar b$. The generated distributions
are parametrized as
\begin{equation}
D_i^{\pi^0} (z,M_{f0}^2) = N_i z^{\alpha_i} (1-z)^{\beta_i}
\label{her2}
\end{equation}
and analyzed using the minimization procedure MINUIT. The coefficients
$N_i$ are constrained by the normalization condition:
\begin{equation}
 \int^1_{{2 m_{\pi} \over M_{f0}}} dz \; D_i (z,M_{f0}^2) =
 {\langle n_{\pi} \rangle}_i,
\end{equation}
where the average values ${\langle n_{\pi} \rangle}_i$ are given
by HERWIG for
each quark flavor, in agreement with the total observed multiplicity
${\langle n_{\pi} \rangle}$. The parameters $N_i,{\alpha_i}$ and
${\beta}_i $ are extracted from the $\pi^0$ inclusive
distribution generated, for each
flavor, in the x range $.025 \leq z_H \leq .95$ and shown in table II.
As can be inferred from this table the statistical error
on the parameters is less than $5\%$.

\indent

As an illustration of the accuracy of the method and also of its
limitations, the $\pi^0$ inclusive cross-section obtained from
eqs.~(\ref{her1}) and~(\ref{her2}), together with
the results of table II
, are compared in figure 2 with the CELLO data
\cite{cello2}
at $\sqrt{S}=35$ GeV. The agreement is reasonable in the range
$z_H \le .5$. So far we have not included the contribution from
the gluon fragmentation function.
Indeed from the analysis of the three jet events it would be possible,
in principle, to extract from HERWIG the appropriate information.
The corresponding accuracy is however unsatisfactory, due to the
limited sensitivity to hard gluon effects in $e^+e^-$ annihilation.

\indent

For this reason we have followed a different approach. To extract
the gluon fragmentation function from HERWIG we have analyzed
the subprocess $ gg \rightarrow gg \rightarrow {\pi^0 + X} $ from
$p \bar p$ annihilation at $M_{f0}=\sqrt s \sim 30$ GeV, in analogy
to the quark case. In order to
eliminate the background from the fragmentation of the spectator partons
we have considered the pions lying only within a cone of
semi aperture $\delta = .35 -.40$ rad
around the direction of the parent gluons emitted at $90 \deg$.
The value of $\delta$ is found by an appropriate angular study of
the generated distribution. With a parametrization of the form
{}~(\ref{her2})  we find the values of the parameters
$N_g,{\alpha_g}$ and ${\beta}_g $ given in table III.
After inclusion of the gluon fragmentation function and use of NLO evolved
fragmentation functions together with NLO terms in the $\pi^0$ inclusive cross
 section the agreement with CELLO data is improved as can be inferred from
figure 3 up to $ z_H\simeq 0.8$

\indent

We compare now our predictions at NLO to experimental data from hadronic
colliders. We first consider the data from CERN ISR \cite{afs,kou}, for
$\sqrt S=63$ GeV, compared in figures 4 and 5 with our predictions for
$\mu=M=M_f=P_t$ and $\mu=M=M_f=P_t/2$ using the quark fragmentation
functions from table II and the two gluon sets from table III, with
$\delta=0.35$ and $\delta=0.40$. The agreement is satisfactory within
the theoretical and experimental uncertainties.

\indent

Let us focus now on the UA2 data at the $Sp\overline{p}S$ collider
\cite{ua2}. We will use two sets of quite precise data, for $P_t\leq
15$ GeV and $\eta\simeq0$ and, for $15\leq P_t \leq 45$ GeV and
$\eta \simeq 1.4$. The comparison with the theoretical predictions is
shown in figures 6 and 7 for $\mu=M=M_f=P_t/2,~P_t$ and for the two gluon
sets of fragmentation functions. The agreement is quite good, and
slightly favors the set corresponding to $\delta=0.35$. The dependence
on the renormalization, factorization and fragmentation scales at NLO will
be discussed later.

\indent

In the next subsection
we will extract the $\pi^0$ fragmentation functions at next to leading
order using two different
hypotheses at the reference scale $M^2_{f0}$ = 2 $\mbox{GeV}^2$.

\subsection{Set I: fragmentation functions with natural scales.}
\indent

For this set, we take $\alpha_s$ as given by
equation~(\ref{apalfa})  and $\Lambda=190$ MeV, corresponding
to the set of structure functions we will use \cite{mt,mrs}.

\subsubsection{Definition}
\indent

We assume for this case an SU(2) symmetry:
\begin{equation}
D^{\pi^0}_{u}(z,M^2_{f0})=D^{\pi^0}_{\bar{u}}(z,M^2_{f0})=
D^{\pi^0}_{d}(z,M^2_{f0})=D^{\pi^0}_{\bar{d}}(z,M^2_{f0})=D_V(z,M^2_{f0})
+ D_S(z,M^2_{f0}) .
\end{equation}
Then, we take
\begin{equation}
D^{\pi^0}_{s}(z,M^2_{f0})=D^{\pi^0}_{\bar{s}}(z,M^2_{f0})=
D^{\pi^0}_{c}(z,M^2_{f0})=D^{\pi^0}_{\bar{c}}(z,M^2_{f0})=D_S(z,M^2_
{f0}),
\end{equation}
and
\begin{equation}
D^{\pi^0}_{g}(z,M^2_{f0})=D_G(z,M^2_{f0}).
\end{equation}
We parametrize the different functions of z as follows
\begin{eqnarray}
D_V(z,M^2_{f0}) & = & N_v \: (1-z)^{\beta_v} \\
D_S(z,M^2_{f0}) & = & N_s \: (1-z)^{\beta_s} \\
D_G(z,M^2_{f0}) & = & N_g \: (1-z)^{\beta_g} .
\end{eqnarray}
At the initial scale $M_{f0}$, we start with four flavors.
The b quark contribution is taken into account
in the evolution.
Fixing the threshold at 4 $m_b^2$, so we have:
\begin{equation}
D_b^{\pi^0}(z,M_f^2) = \left\{ \begin{array}{ll}
             0 & \mbox{if $M_f^2<4m_b^2$} \\
             N_s \: (1-z)^{\beta_s} & \mbox{if $M_f^2=4m_b^2$}
             \end{array}
             \right.
\end{equation}
So we are left with six parameters to be determined with
the help of experimental data. \\

\subsubsection{Choice of the scale}
\indent

We use the standard approach to fix all the scales to the
same value which is some natural scale of the problem.
More precisely, for $e^+ e^-$ collisions, we take $\mu=M_f=
\sqrt S $ whereas for p p collisions, we
set the three scales equal and
proportional to the transverse momentum of the $\pi^0$:
\[ \mu=M=M_f= c P_t\]
where c is a constant to be fixed by the fit to experimental data .

\subsubsection{Results for set I}
\indent

First of all, for $e^+ e^-$ collisions, we limit ourselves to
a $\pi^0$ energy greater than 2 GeV because we don't trust
perturbation theory for low $\pi^0$ energies.
Therefore
for $\sqrt S \simeq 30$ GeV, we will only use z values greater than
0.1. As it can be inferred from eqs (25-27) we have not used a factor
 $z^{\alpha}$
in the input parametrizations since in
this z range it does not improve the
fit but only leads to correlations. With  six parameters, a big
correlation still occurs between $N_v$ and $\beta_v$,
so we fix $\beta_v=1$. Then $N_s$, $N_g$ and $\beta_g$ remain
slightly correlated.
A good fit to CELLO~\cite{cello2}, TASSO~\cite{tasso},
TPC~\cite{tpc} and JADE~\cite{jade}
data leading to a $\chi^2 = 26.3$ for 29 points is obtained for values
of the parameters given in table IV (systematic errors have been added
in quadrature to statistical errors).

\indent

Using set I of fragmentation functions we will now evaluate the NLO
cross-sections for inclusive $\pi^0$ production in hadronic
collisions and compare them to experimental data from low center of mass
energies up to the CERN collider one.
Here, the situation is less clear. First, if we keep constant
the value of the parameter
c  it is impossible to obtain a good fit in the whole energy domain.
For example, setting $c \simeq 1.5$, the ISR data can be
described but the theoretical predictions are by far too low for WA70 and E706
and too high for UA2.
A simple solution to this problem is to allow c to vary
 with the hadronic kinematical variables, in particular $\sqrt S$.
A correct description of the data requires  $c \simeq 0.39$ for WA70
\cite{wa70}
(see figure 8a),
$c \simeq 0.5$ for E706  \cite{e706}(see figure 8b),
$c \simeq 1.5$ for ISR experiment \cite{afs,kou}  (see figure 8c)
and $c \simeq 5.5$
for UA2 \cite{ua2} (see figure 8d).
In particular for the ISR
energy range, the data from AFS collaboration \cite{afs} are
marginally consistent with those of reference \cite{kou}
since the transverse momentum dependence in the two experiments
is different. Therefore it is very difficult to describe
both ISR data with high precision. We get rather good fits of data of
Kourkoumelis et al. \cite{kou} with
$\chi^2 = 20.6$ for $14$ points using  $\mu =M =M_f = 1.3 P_t$ and of
the AFS collaboration \cite{afs} with
$\chi^2 = 12.2$ for $11$ points using  $\mu =M =M_f = 1.6 P_t$.
Notice that the slope of the UA2 data is not correctly reproduced,
with a $\chi^2 =50.2$ for $11$ points. The $\chi^2$ have been
calculated with statistical errors, allowing the overall
normalization to vary within the systematic error.
\indent

A comment is in order here. The approach followed so far
is rather simple.
When the energy grows up the scales
needed to describe data have also to increase. As stated above
an acceptable fit of UA2 data  \cite{ua2} in the
forward direction can be obtained for the choice of scales
$\mu =M =M_f = 5.5P_t$ which is a priori a large scale.
The
compensation occurring between the  leading and next-to-leading terms
concerning the scale dependence is much more effective at high energies.
At low energy, since we prevent the scale to be less than $M_{f0}=\sqrt 2$ GeV,
this compensation does not occur and the behavior of the leading and
next-to-leading cross-sections is quite the same. In other words,
we are not in a good region to perform perturbation theory.

\indent
This approach might be criticized.
 Indeed it is not very predictive, since the scales
change with the energy. In other words one adds a new parameter which
acts as an overall normalization for each experiment. Notice that
the normalization of the glue fragmentation function
$N_g$ is strongly correlated to the choice made for the scale. More
precisely, we could perfectly find a value for $N_g$ which describes the
UA2 data with $c=0.5$. But in this case we couldn't describe the other
data at lower energies.

\indent

\subsection{Set II: fragmentation functions with optimized scales.}
\indent

For this set, we take the numerical solution of equation~(\ref{trualfa})
for $\alpha_s$ and $\Lambda=230$ MeV, since we will use the ABFOW set of
structure functions \cite{abf} .

\subsubsection{Definition}
\indent

We assume also for this case an SU(2) symmetry:
\begin{equation}
D^{\pi^0}_{u}(z,M^2_{f0})=D^{\pi^0}_{\bar{u}}(z,M^2_{f0})=
D^{\pi^0}_{d}(z,M^2_{f0})=D^{\pi^0}_{\bar{d}}(z,M^2_{f0})=
D_u(z,M^2_{f0}).
\end{equation}
Then we take:
\begin{eqnarray}
D^{\pi^0}_{s}(z,M^2_{f0})=D^{\pi^0}_{\bar{s}}(z,M^2_{f0}) & = &
D_s(z,M^2_{f0}),  \\
D^{\pi^0}_{c}(z,M^2_{f0})=D^{\pi^0}_{\bar{c}}(z,M^2_{f0}) & = &
D_c(z,M^2_{f0}),
\end{eqnarray}
and
\begin{equation}
D^{\pi^0}_{g}(z,M^2_{f0})=D_g(z,M^2_{f0}).
\end{equation}
We parametrize these different functions of z in the following
way:
\begin{eqnarray}
D_u(z,M^2_{f0}) & = & N_u \: z^{-1} \: (1-z)^{\beta_u} \\
D_s(z,M^2_{f0}) & = & N_s \: z^{-1} \: (1-z)^{\beta_s} \\
D_c(z,M^2_{f0}) & = & N_c \: z^{-1} \: (1-z)^{\beta_c} \\
D_g(z,M^2_{f0}) & = & N_g \: z^{-1} \: (1-z)^{\beta_g} .
\end{eqnarray}
So we are left with eight parameters to be determined with
the help of experimental data.
Since we will use the optimized procedure for the determination of the
scales, it is much simpler not to change the number of flavors.
So, in this case, we will neglect the b contribution.
This assumption is motivated by the fact that
$\sigma(e^+ \: e^- \rightarrow \gamma^*\rightarrow b \: \bar{b})
= 1/4 \: \sigma(e^+ \: e^- \rightarrow \gamma^* \rightarrow c \:
\bar{c})$
and in $p \: p$ collision the b production
is suppressed due to the weak b content of the proton.
\par
A few remarks are in order here.
As in the case of set I, the non singlet part $D^-_i$
is always zero due to our assumptions.
We did not take $D_s^{\pi^0} = D_c^{\pi^0}$ because in this
case the sum over the four flavors of $D^+_i$ weighted
by the square electric charge is zero:
\[ \sum_{i=u,d,s,c} e_i^2 \left( D^+_i(z,M^2)+D^+_{\bar{\imath}}
(z,M^2) \right) = 0 .  \]
So, there is no non-singlet contribution to the cross-section.
Therefore we could parametrize directly the singlet and the glue with four
parameters only. The $e^+ e^-$ data could be
correctly described, but the glue is very constrained and
it will not be possible
to fit hadronic data in the whole energy range.

\subsubsection{Choice of the scale}
\indent

For set II, we use  optimized
scales according to the procedure of Politzer and Stevenson \cite{steve}.
Concerning  $e^+ e^-$ collisions,
our approach is the following.
Firstly since the scale $\mu$ does not appear at
lowest order, we cannot optimize with respect to it. Therefore
we set $\mu = M_f$ and
perform an optimization only with respect to the scale $M_f$.
Therefore, a priori, our optimized scale
depends on the choice made for the input fragmentation functions.
We have not found a way to get rid from this sensitivity.
In practice, the optimized point changes slowly when the
input is modified and in addition, since we are in a stable region, it does not
matter if we are not exactly on the optimized point.
The optimized scale $M_f^{opt}$ is of order of $\sqrt S /8$ varying
slowly with z. Furthermore, we find no optimization scale for $z \leq
.03$ for $\sqrt S=35$ GeV, $z \leq .05$ for $\sqrt S = 29$ GeV and
$z \leq .1$
for $\sqrt S = 22$ GeV. For lower values of $\sqrt S$, it is
not possible to optimize.

\indent

 We also use an optimization procedure for hadronic collisions.
 So we require
 that:
\begin{eqnarray}
\frac{\partial}{\partial \ln(\mu^2/ \Lambda^2)} \; E_{\pi^0}
\frac{d \sigma_{p+p \rightarrow \pi^0}}{d^3 \vec{P}_{\pi^0}} & = &
0 \\
\frac{\partial}{\partial \ln(M^2/ \Lambda^2)} \; E_{\pi^0}
\frac{d \sigma_{p+p \rightarrow \pi^0}}{d^3 \vec{P}_{\pi^0}} & = &
0 \label{derivm} \\
\frac{\partial}{\partial \ln(M_f^2/ \Lambda^2)} \; E_{\pi^0}
\frac{d \sigma_{p+p \rightarrow \pi^0}}{d^3 \vec{P}_{\pi^0}} & = &
0 \label{derivmf}.
\end{eqnarray}
The first equation can be computed analytically:
\begin{eqnarray}
\frac{\partial}{\partial \ln(\mu^2/ \Lambda^2)} \; E_{\pi^0}
\frac{d \sigma_{p+p \rightarrow \pi^0}}{d^3 \vec{P}_{\pi^0}} & = &
- \alpha_s^4(\mu^2) b \left\{ 2 b' A + 3 (1+b' \alpha_s(\mu^2))
\left[ 2 b A \ln\!\left(\frac{\mu^2}{\Lambda^2}\right) \right.
\right. \nonumber \\
& & \mbox{} \left. \left. + B \ln\!\left( \frac{M^2}{\Lambda^2} \right)
+ C \ln\!\left(\frac{M_f^2}{\Lambda^2}\right)
+ D \right] \right\} \label{derivmu}
\end{eqnarray}
having used
\begin{equation}
\frac{\partial \alpha_s(\mu^2)}{\partial \ln(\mu^2/\Lambda^2)}  =
- b \; \alpha_s^2(\mu^2) \; (1+ b' \alpha_s(\mu^2)) .
\end{equation}
Note that terms of order of $\alpha_s^3$ have been cancelled as it should
be. Now, we determine the scale $\mu$ in order to cancel the right-hand side of
eq~(\ref{derivmu}). This ensures us that the corrective
term K will be negative with a magnitude of roughly 10 \% of the lowest order.
Then we compute numerically the value of the
scales $M$ and $M_f$ which  have to fulfill the
equations~(\ref{derivm})~(\ref{derivmf}), the scale $\mu$ being now a
function of $M$, $M_f$. We require that the factorization scales must
be greater than $\sqrt 2$ GeV and that the renormalization scale is such
that the running coupling constant $\alpha_s$ is less than .34.
With these constraints it will be impossible to optimize in low $P_t$ range.
More precisely, for low center of mass energies ($\sqrt S \leq 63$ GeV),
the optimization is not possible for $P_t \leq 5$ GeV. Therefore these regions
are not appropriate to apply an optimization procedure.

\subsubsection{Results for set II}
\indent

First we freeze $\beta_s$, $\beta_c$ and $\beta_g$ according to the counting
rules.
There are still too many parameters, so we fix $N_g$ and fit to $e^+ e^-$ data
 with four parameters $N_u$, $\beta_u$, $N_s$ and $N_c$. The fragmentation
functions
extracted are then used to evaluate  hadronic cross sections. Then we vary
$N_g$ refitting  $e^+ e^-$ data and apply the new input
 to pp data.This procedure is repeated until a reasonable description of
hadronic data is reached. Good fits of $e^+ e^-$ data
(CELLO~\cite{cello1,cello2}, TASSO~\cite{tasso}, TPC~\cite{tpc} and
JADE~\cite{jade})
leading to a $\chi^2 \simeq 1$ per d.o.f. are obtained for the two sets
- hereafter denoted as set IIa and set IIb -
displayed in Table V and Table VI (see figures 9 and 10 using set IIb).
The two sets differ mainly for the gluon
normalization. As can be seen from inspection of
figures 11a, 11b, 12a and 12b a rather good fit of the latest UA2 data at
$\sqrt S =630$ GeV \cite{ua2},
AFS \cite{afs} and Kourkoumelis et al data \cite{kou}  can be obtained
leading to a
$\chi^2 \simeq 50$ for 31 points. Kourkoumelis et al. data favor
the set characterized by the largest glue (set IIb) whereas UA2 data are better
fitted by the other set (set IIa).
Notice that we have taken into account the systematic errors of the
data which affect the overall
normalization. The $\chi^2$ are $3.46$ ($4.28$) for the 11 AFS points,
$31.54$ ($23.52$) for the 9 Kourkoumelis et al. points and
$14.91$ ($20.00$) for the 11 UA2 points with the  parameters of set
IIa (IIb).
 Inside the
systematic errors we can also describe UA2 data at $\sqrt S = 540$ GeV
and $\eta = 1.4$. On the other hand we are not able
to describe  WA70 and E706 data with the values of $N_g$ found
before.
This is not very surprising since the corrective term is found to be huge, and
 although we can find an optimization point this is not very stable suggesting
that we are not in the appropriate region to trust perturbation theory.

\section{Predictions at future hadron colliders.}
\indent

As we have seen present data do not allow to extract the $\pi^0$
fragmentation functions unequivocally.
To this aim the forthcoming information
from $ep$ HERA collider should be very helpful. With these limitations
we will now estimate the $\pi^0$ rates at LHC
using the various sets of fragmentation functions previously derived.

\indent

Let us consider first set I of fragmentation functions.
In order to describe hadronic data we had to increase the scales
$\mu =M =M_f$ from $P_t \over 2$ at
$\sqrt S \simeq 20$ GeV up to $5 P_t$ at $\sqrt S =630$ GeV.
An extrapolation
to LHC energy would lead to $\mu =M =M_f \simeq 50 P_t$ which seems
by far an unnatural scale. To estimate the sensitivity to scales
we show in figure 13 the ratio of cross sections at LHC for the two scales
$50 P_t$ and $P_t$ at $\eta=0$. As can be inferred from the figure the rates
differ by
at most a factor of two. To estimate the uncertainty due to structure functions
we have taken the set of structure functions of
HMRS \cite{mrs} using the $\overline{MS}$ scheme and the set of Morfing-Tung
\cite{mt} using the DIS scheme. The predictions differ by at most $20\%$.
Similarly the ratio of predictions using set II is displayed in figure 14.

\indent

Finally the ratio of the two predictions from the HERWIG fragmentation
functions,~for $\delta$ =0.35,~0.40,~evolved to NLO accuracy as discussed
in section 3.2,~are displayed in figure 15.

\indent

The situation is summarized in figure 16
where we show the absolute rates at LHC for $\eta=0$ from the most plausible
sets in the three
approaches. This gives an estimate of the theoretical uncertainty which is
of the order of a factor two.
 The uncertainty
on structure functions is marginal compared to the poor determination of
fragmentation functions.
\indent

To show the stability of the NLO corrections
 we display  the cross
section as a function of the scales $\mu$ and $M=M_f$ compared to the LO result
for $P_t=50$ GeV (figures 17) and for $P_t=200$ GeV (figures 18). We vary the
scales between $P_t/5$ and $5P_t$. The NLO cross sections exhibit a saddle
point
whereas the LO cross sections decrease monotonically when the scales increase.
\indent

The uncertainty due to factorisation scheme,~especially coming from
fragmentation
functions is expected to be tiny for the two following reasons. Firstly the
evaluation done for one jet inclusive cross section has shown[2] that at
collider
energies its magnitude is of the order of $5\%$ -if done correctly - and
we can reasonably expect a same order of magnitude for one hadron inclusive
cross section. Secondly a precise estimate doesn't seem mandatory compared to
the
large uncertainty coming from fragmentation functions.
\vfill\eject
\par\noindent
\section{Conclusions}
\indent

We have performed a  complete next to leading order analysis of inclusive
$\pi^0$ production from $e^+e^-$ and hadronic data.
For the first time an attempt
to extract sets of $\pi^0$ fragmentation functions at NLO has
been performed. The
present quality of data does not allow us to derive a unique set
fitting all experimental data. For this purpose more
accurate measurements from hadronic colliders
in the large $P_t$ domain, from $e^+e^-$ colliders in the large $z_H$ domain
 as well as complementary information from $ep$
collisions will be very helpful. The theoretical uncertainties
are mainly due to the poorly determined fragmentation functions. Nevertheless
the absolute rates at future
colliders like LHC and SSC can be predicted within a factor of two. This will
certainly be of help for neutral background rejection at supercolliders.
\vskip 40pt
\par\noindent
{\bf {Note added in proof}}
\par
After completion of this work the paper "Higher order QCD corrections to
inclusive particle production in $p\bar p$ collision" by F.~M.~Borzumati et
al. \cite{franc} has appeared, where the $\pi^0$ inclusive production has been
discussed,~using the old LO fragmentation functions of ref.[8] and the NLO
results of our group \cite{acgg}.
\vskip 40pt
\par\noindent
{\bf Acknowledgements}
\par
We would like to thank P. Nason for providing  us his fortran
code for b fragmentation at next to leading order.
We are greatly indebted to G. Marchesini for enlightening discussions
concerning HERWIG and also to P. Aurenche for advice
on optimization procedure. We acknowledge discussions on the ISR data
with C. Kourkoumelis.
\vfill\eject
\par\noindent
{\bf Table captions}
\vskip 24 pt

\begin{itemize}
\item{Table I:} average fraction of energy of the fragmenting parton (see
eq.18)
 from pp collisions.
\item{Table II:} parameters of the quark fragmentation functions (see eq.20) as
 obtained from HERWIG in $e^+e^-$ annihilation at $M_0=30$ GeV.
\item{Table III:} parameters of the gluon fragmentation functions (see eq.20)
as
 obtained from HERWIG  at $M_0=30$ GeV, with two hypotheses on the angle
$\delta$ (see text).
\item{Table IV:} parameters for the $\pi^0$ fragmentation functions (set I)
obtained from $e^+e^-$ annihilation.
\item{Table V:} parameters for the $\pi^0$ fragmentation functions (set IIa)
obtained from optimization procedure in $e^+e^-$ annihilation.
\item{Table VI.} Parameters for the $\pi^0$ fragmentation functions (set IIb)
obtained from optimization procedure in $e^+e^-$ annihilation.
\end{itemize}

\vfill\eject

\vskip 5pt
$$
\begin{tabular}{|c|c||c|c||c|c|}
\hline
\multicolumn{2}{|c||}{$\sqrt S = 23$ GeV and $\eta=0.$}
&\multicolumn{2}{c||}{$\sqrt S = 63$ GeV and $\eta=0.$}
&\multicolumn{2}{c|}{$\sqrt S = 630$ GeV and $\eta=1.4$} \\
\hline
\multicolumn{1}{|c|}{$P_t^{\pi^0}$} & \multicolumn{1}{c||}{$<z>$}
&\multicolumn{1}{c|}{$P_t^{\pi^0}$} & \multicolumn{1}{c||}{$<z>$}
&\multicolumn{1}{c|}{$P_t^{\pi^0}$} & \multicolumn{1}{c|}{$<z>$} \\
\hline
4.11 & 0.81 & 5.25 & 0.67 & 13 & 0.55 \\ \hline
4.61 & 0.82 & 6.73 & 0.70 & 21 & 0.60 \\ \hline
5.69 & 0.86 & 8.23 & 0.73 & 29.8 & 0.65 \\ \hline
6.69 & 0.89 & 10.4 & 0.77 & 43.7 & 0.74 \\ \hline
\end{tabular}
$$
\centerline{\bf{ Table I}}\rm{
$$
\vbox{\offinterlineskip
\hrule
\halign{&\vrule#&
\strut\quad\hfil#\quad\cr
height4pt&\omit&&\omit&&\omit&&\omit&&\omit&\cr
& \it{Process}\hfill&&\hfill $\alpha$\hfill&&\hfill $\beta$\hfill&&\hfill
$N_q$\hfill&&\hfill$<n_{\pi}>$\hfill&\cr
height4pt&\omit&&\omit&&\omit&&\omit&&\omit&\cr
\noalign{\hrule}
height4pt&\omit&&\omit&&\omit&&\omit&&\omit&\cr
&\hfill $e^+e^-\to u\bar u$ \hfill&& \hfill $-0.95\pm 0.02$\hfill&&
 \hfill $3.67\pm 0.19$
\hfill&& \hfill $1.20$\hfill&&\hfill$2.95$\hfill& \cr
height4pt&\omit&&\omit&&\omit&&\omit&&\omit&\cr
&\hfill $e^+e^-\to d \bar d$ \hfill&& \hfill $-0.95\pm 0.02$\hfill&&
\hfill $3.67\pm 0.15$
\hfill&& \hfill $1.24$ \hfill&&\hfill$2.87$\hfill& \cr
height4pt&\omit&&\omit&&\omit&&\omit&&\omit&\cr
&\hfill $e^+e^-\to s \bar s$ \hfill&& \hfill $-0.88\pm 0.02$ \hfill&&
\hfill $5.32\pm 0.23$
\hfill&& \hfill $1.68$\hfill&&\hfill$2.73$\hfill& \cr
height4pt&\omit&&\omit&&\omit&&\omit&&\omit&\cr
&\hfill $e^+e^-\to c \bar c$ \hfill&& \hfill $-0.82\pm 0.02$ \hfill&&
\hfill $8.02\pm 0.24$
\hfill&& \hfill $3.09$\hfill&&\hfill$3.42$\hfill& \cr
height4pt&\omit&&\omit&&\omit&&\omit&&\omit&\cr
&\hfill $e^+e^-\to b \bar b$ \hfill&& \hfill $-0.95\pm 0.02$ \hfill&&
\hfill $10.94\pm 0.29$
\hfill&& \hfill $2.92$\hfill& &\hfill$4.20$\hfill&\cr
height4pt&\omit&&\omit&&\omit&&\omit&&\omit&\cr}
\hrule}
$$
\centerline{\bf{ Table II}}\rm{
$$
\vbox{\offinterlineskip
\hrule
\halign{&\vrule#&
\strut\quad\hfil#\quad\cr
height4pt&\omit&&\omit&&\omit&&\omit&&\omit&\cr
& $\delta$\hfill&&\hfill $\alpha$\hfill&&\hfill $\beta$\hfill&&\hfill
$N_g$\hfill&&\hfill$<n_{\pi}>$\hfill&\cr
height4pt&\omit&&\omit&&\omit&&\omit&&\omit&\cr
\noalign{\hrule}
height4pt&\omit&&\omit&&\omit&&\omit&&\omit&\cr
&\hfill $0.35~rad$ \hfill&& \hfill $-0.28\pm 0.04$\hfill&& \hfill
$6.71\pm 0.39$
\hfill&& \hfill $14.49$\hfill&&\hfill$3.65$\hfill& \cr
height4pt&\omit&&\omit&&\omit&&\omit&&\omit&\cr
&\hfill $0.4~rad$ \hfill&& \hfill $-0.37\pm 0.04$\hfill&& \hfill
$5.79\pm 0.36$
\hfill&& \hfill $12.93$ \hfill&&\hfill$4.55$\hfill& \cr
height4pt&\omit&&\omit&&\omit&&\omit&&\omit&\cr}
\hrule}
$$
\centerline{\bf{Table III}}\rm{

$$
\vbox{\offinterlineskip
\hrule
\halign{&\vrule#&
\strut\quad\hfil#\quad\cr
height4pt&\omit&&\omit&&\omit&&\omit&\cr
& \it{Parton}\hfill&&\hfill $\alpha_i$\hfill&&\hfill $\beta_i$\hfill&&\hfill
$N_i$\hfill&\cr
height4pt&\omit&&\omit&&\omit&&\omit&&\omit&\cr
\noalign{\hrule}
height4pt&\omit&&\omit&&\omit&&\omit&\cr
&\hfill $valence $ \hfill&& \hfill $ 0. $\hfill&& \hfill
$ 1. $
\hfill&& \hfill $ 0.19 $\hfill& \cr
height4pt&\omit&&\omit&&\omit&&\omit&\cr
&\hfill $ sea  $ \hfill&& \hfill $ 0. $\hfill&&
\hfill $ 5.2 $
\hfill&& \hfill $ 3.5 $\hfill& \cr
height4pt&\omit&&\omit&&\omit&&\omit&\cr
&\hfill $gluon$ \hfill&& \hfill $ 0. $ \hfill&&
\hfill $ 2.03 $
\hfill&& \hfill $ 4.9 $\hfill& \cr
height4pt&\omit&&\omit&&\omit&&\omit&&\omit&\cr}
\hrule}
$$
\centerline{\bf{ Table IV}}\rm{

$$
\vbox{\offinterlineskip
\hrule
\halign{&\vrule#&
\strut\quad\hfil#\quad\cr
height4pt&\omit&&\omit&&\omit&&\omit&\cr
& \it{Parton}\hfill&&\hfill $\alpha_i$\hfill&&\hfill $\beta_i$\hfill&&\hfill
$N_i$\hfill&\cr
height4pt&\omit&&\omit&&\omit&&\omit&&\omit&\cr
\noalign{\hrule}
height4pt&\omit&&\omit&&\omit&&\omit&\cr
&\hfill $ up  $ \hfill&& \hfill $ -1. $\hfill&& \hfill
$ 0.94 $
\hfill&& \hfill $ 0.11 $\hfill& \cr
height4pt&\omit&&\omit&&\omit&&\omit&\cr
&\hfill $ strange  $ \hfill&& \hfill $ -1. $\hfill&&
\hfill $ 3.0 $
\hfill&& \hfill $0.55 $\hfill& \cr
height4pt&\omit&&\omit&&\omit&&\omit&\cr
&\hfill $charm$ \hfill&& \hfill $ -1. $ \hfill&&
\hfill $ 4. $
\hfill&& \hfill $ 2.7 $\hfill& \cr
height4pt&\omit&&\omit&&\omit&&\omit&\cr
&\hfill $gluon$ \hfill&& \hfill $ -1. $ \hfill&&
\hfill $ 2. $
\hfill&& \hfill $ 0.55 $\hfill& \cr
height4pt&\omit&&\omit&&\omit&&\omit&&\omit&\cr}
\hrule}
$$
\centerline{\bf{ Table V}}\rm{

$$
\vbox{\offinterlineskip
\hrule
\halign{&\vrule#&
\strut\quad\hfil#\quad\cr
height4pt&\omit&&\omit&&\omit&&\omit&\cr
& \it{Parton}\hfill&&\hfill $\alpha_i$\hfill&&\hfill $\beta_i$\hfill&&\hfill
$N_i$\hfill&\cr
height4pt&\omit&&\omit&&\omit&&\omit&&\omit&\cr
\noalign{\hrule}
height4pt&\omit&&\omit&&\omit&&\omit&\cr
&\hfill $ up  $ \hfill&& \hfill $ -1. $\hfill&& \hfill
$ 1.11 $
\hfill&& \hfill $ 0.15 $\hfill& \cr
height4pt&\omit&&\omit&&\omit&&\omit&\cr
&\hfill $ strange  $ \hfill&& \hfill $ -1. $\hfill&&
\hfill $ 3.0 $
\hfill&& \hfill $ 0.18 $\hfill& \cr
height4pt&\omit&&\omit&&\omit&&\omit&\cr
&\hfill $charm$ \hfill&& \hfill $ -1. $ \hfill&&
\hfill $ 4. $
\hfill&& \hfill $ 2.5 $\hfill& \cr
height4pt&\omit&&\omit&&\omit&&\omit&\cr
&\hfill $gluon$ \hfill&& \hfill $ -1. $ \hfill&&
\hfill $ 2. $
\hfill&& \hfill $ 0.75 $\hfill& \cr
height4pt&\omit&&\omit&&\omit&&\omit&&\omit&\cr}
\hrule}
$$
\centerline{\bf{ Table VI}}\rm{
\vfill\eject
{\bf Figure Captions}
\vskip 24 pt

\begin{itemize}
\item{Fig. 1:} relative inclusive partonic production in hadronic collisions
as a function of the partonic transverse momentum $P_{tl}$ at various energies:
WA70 experiment(fig 1a),~ISR experiments(fig 1b),UA2(fig 1c) and LHC(fig 1d).
The
various curves refer to: $pp \rightarrow u+d$ (full line), $pp \rightarrow
 s+c$ (dashed line) and  $pp \rightarrow g$ (dot-dashed line).
\item{Fig. 2:} LO inclusive $\pi^0$ production in
$e^+e^-$ annihilation with the
quark fragmentation functions extracted from HERWIG at $\sqrt S= 30$ GeV,
compared with experimental data at  $\sqrt S= 35$ GeV.
\item{Fig. 3:} NLO inclusive $\pi^0$ production in $e^+e^-$
 annihilation with the
quark and gluon fragmentation functions extracted from HERWIG and evolved
at $\sqrt S= 35$ GeV, compared with data.
\item{Figs. 4:} NLO inclusive $\pi^0$ production in $pp$ collisions at ISR
energies for $\mu=M=M_f= P_t,P_t/2 $ for $\delta =0.35$ (see text). The
fragmentation functions are from HERWIG.
\item{Figs. 5:} same as fig 4, for $\delta =0.40$.
\item{Figs. 6:}  NLO inclusive $\pi^0$ production in $p \bar p$ collisions at
 $Sp \bar p S $ energies
 for $\mu=M=M_f= P_t,P_t/2 $ for $\delta =0.35$  and $\delta =0.40$,at
$\sqrt S=540$ GeV and $\eta=0$. The
fragmentation functions are from HERWIG.
\item{Figs. 7:}  same as fig 6 at $\sqrt S=630$ GeV and $\eta=1.4$.
\item{Figs. 8:}  NLO inclusive $\pi^0$ production in hadronic collisions with
set I fragmentation functions (see text) for various energies. The scales
$\mu=M=M_f=c P_t $ are indicated explicitely.
\item{Fig. 9:}  NLO inclusive $\pi^0$ production
in $e^+e^-$ annihilation with
set II of fragmentation functions compared to CELLO, TASSO and JADE data
. The gluon parameter $N_g$ takes the value
$N_g=0.75$. Data and theory have been
multiplied by 0.1 at $\sqrt{S}=22$ GeV.

\item{Fig. 10:}  Same as fig 9 for TPC data.
\item{Figs. 11:} NLO inclusive $\pi^0$ production in $pp$ collisions with set
II
fragmentation functions (see text) at ISR energies (squares correspond to AFS
data whereas circles correspond to Kourkoumelis et al data).
 The gluon parameter $N_g$ takes the value $N_g=0.55$ (fig 11a) and $N_g=0.75$
(fig 11b).
\item{Figs. 12:} same as fig 11 at $Sp \bar p S $ energies.
\item{Fig. 13:} ratio of inclusive $\pi^0$ cross sections predicted at LHC for
$\eta=0$
using set I of fragmentation functions with $\mu=M=M_f=50 P_t $ over
$\mu=M=M_f=P_t $.
\item{Fig. 14:} ratio of inclusive $\pi^0$ cross sections predicted at LHC
for $\eta=0$ using set IIb of fragmentation functions over set IIa .
\item{Figs. 15:} ratio of inclusive $\pi^0$ cross sections predicted at LHC
for $\eta=0$ using fragmentation functions from HERWIG.
\item{Fig. 16:}  inclusive $\pi^0$ cross sections predicted at LHC for $\eta=0$
using various fragmentation functions : HERWIG with $\delta=0.35$ (full line),
set I with $\mu=M=M_f=50 P_t $ (dot-dashed curve) and set II with  $N_g=0.75$ (
dashed curve).
\item{Figs. 17:} inclusive $\pi^0$ cross sections at LHC in pb using set I of
fragmentation functions as a function of the scales $\mu$ and $M=M_f$ for
 $P_t=50$ GeV and $\eta=0$. LO prediction : fig 17a . NLO prediction : fig 17b.
\item{Figs. 18:} same as fig 17 for $P_t=200$ GeV and $\eta=0$.
\end{itemize}

\vfill\eject

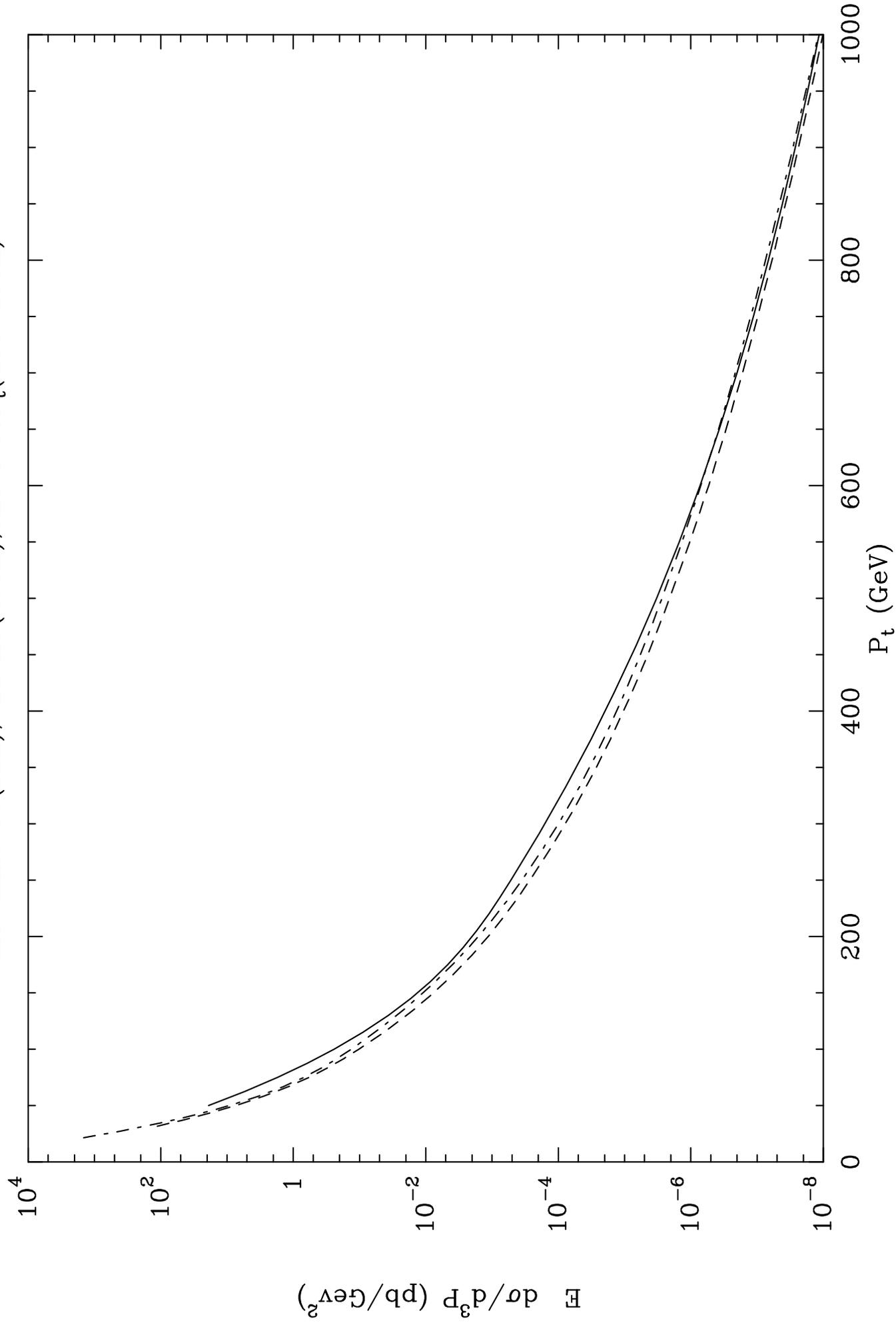
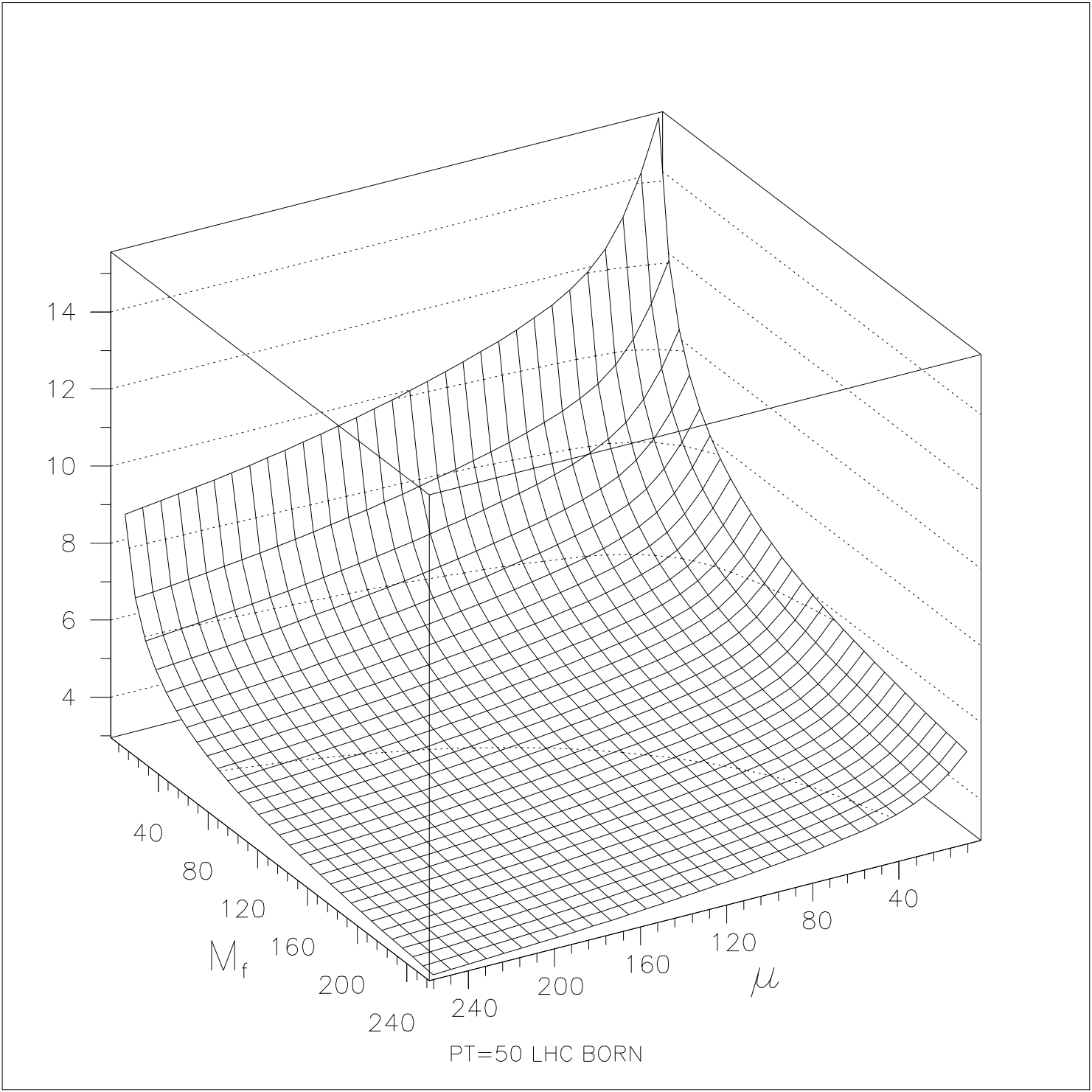
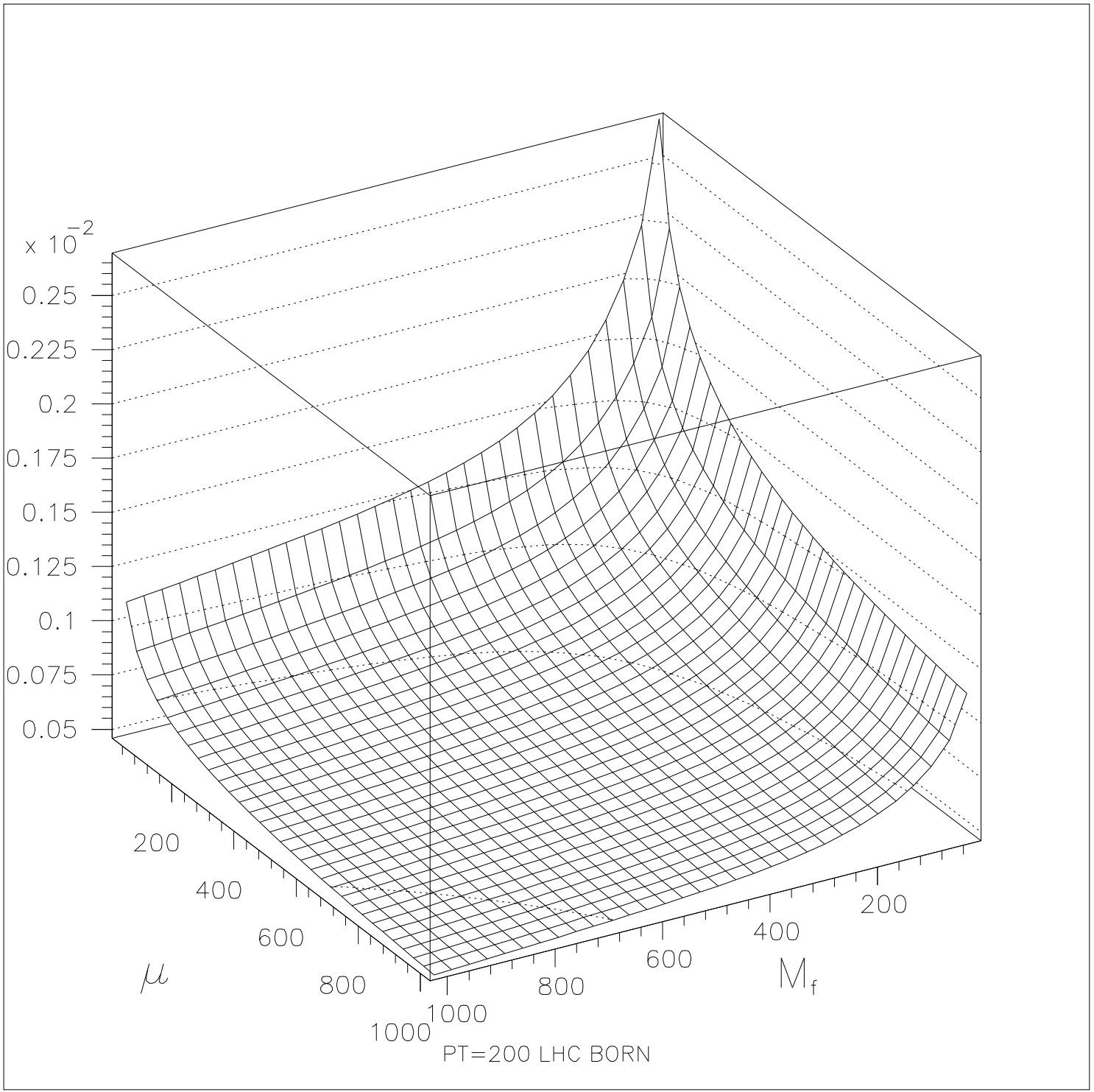
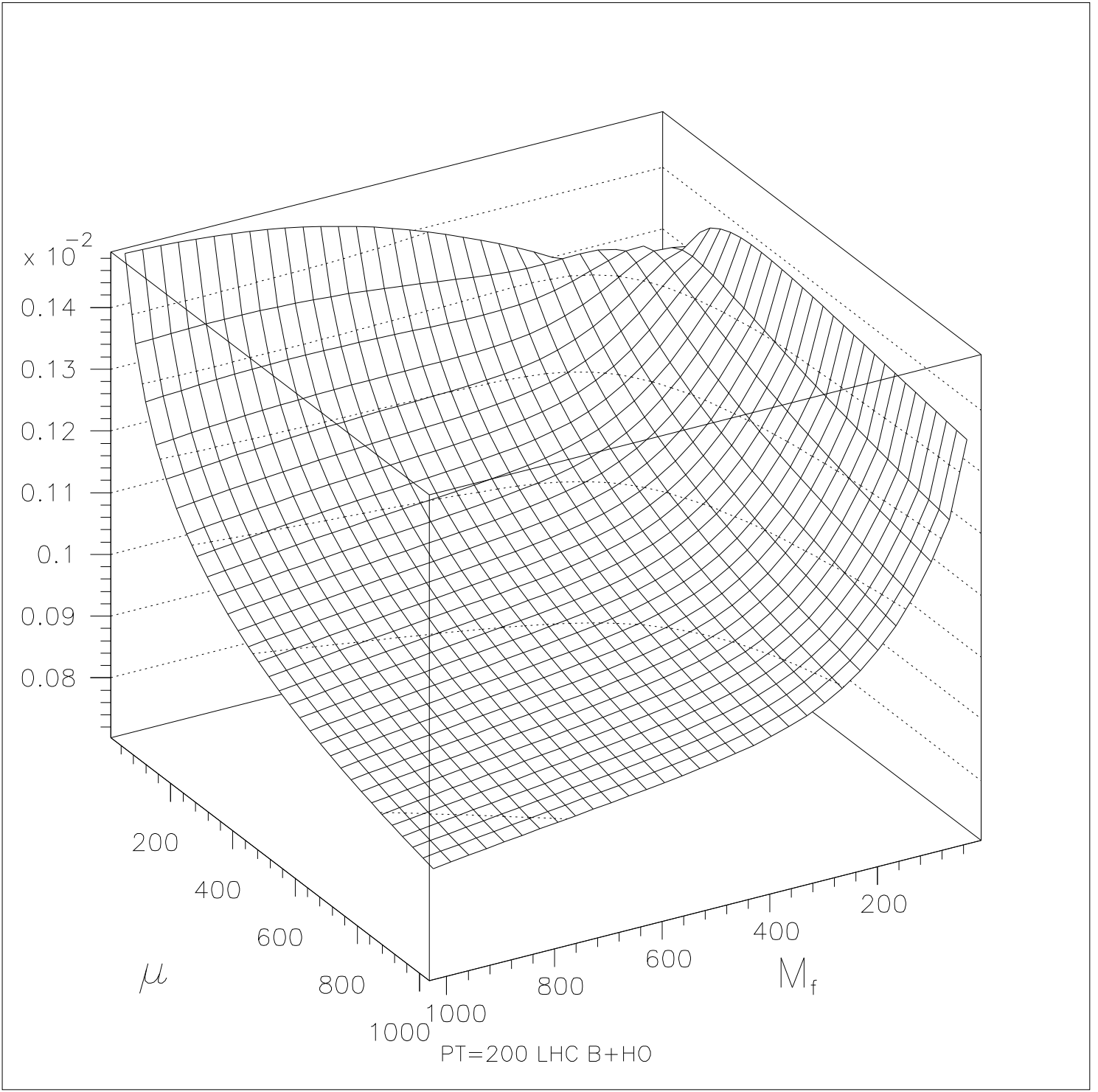
\end{document}